\documentclass[12pt]{article}
\usepackage{epsf}
\usepackage{color}
\usepackage{amsmath,amssymb,color,graphics,amscd,amsfonts}

\begin{document}

\begin{flushright} 
 Febrero, 2010  \\
 FPAUO-10/01 \\
\end{flushright} 

\vspace{0.1cm}

\begin{Large}
\vspace{1cm}
\begin{center}
{\bf Quarter BPS classified by Brauer algebra} \\ 
\end{center}
\end{Large}

\vspace{1.4cm}

\begin{center}
{\large Yusuke Kimura}   \\ 

\vspace{0.7cm} 
{\sf kimurayusuke@uniovi.es}
\vspace{0.8cm} 

{\it Departamento de Fisica, 
Universidad de Oviedo, \\
Avda. Calvo Sotelo 18, 33007 Oviedo, Spain}

\end{center}

\vspace{1cm}

\begin{abstract}
\noindent 
\end{abstract}

We analyse the one-loop dilatation operator with the help of 
the Brauer algebra. 
We find some BPS operators in ${\cal N}=4$ SYM, 
which are labelled by irreducible representations of 
the Brauer algebra. 
Some of them are quarter BPS operators. 
The result includes 
full non-planar corrections. 
Our construction and proof are based on simple algebraic arguments 
and are carried out for any number of fields.

\newpage 

\section{Introduction and Summary}

\hspace{0.46cm}

The study of conformal field theories has been an important topic 
especially in understanding string theory. 
In particular ${\cal N}=4$ Super Yang-Mills theory has been extensively 
investigated in the context of the AdS/CFT correspondence. 
Concretely a large number of progress has been made 
in the computation of 
scaling dimensions of local operators. 
In general, 
to determine the scaling dimensions 
is not an easy task because of the operator mixing problem 
\cite{9911222,0205321,0206020,0307081,0109064}.  
In spite of this problem, 
if we restrict our attention to the planar theory, 
the operator mixing problem can be simplified 
because 
the dilatation operator may be identified with the 
integrable spin-chain system \cite{0212208,0303060,0307042}. 
In contrast, our understanding of the full theory including non-planar corrections 
is not still enough to handle the problem.  
Operators with definite scaling dimensions can be 
linear combinations of single traces and multi-traces. 
To tackle the mixing problem including non-planar corrections, 
we would need to find a good way to organise gauge invariant operators of 
both single trace and multi-trace.

\vspace{0.4cm}

In our previous paper \cite{0709.2158}, 
we have proposed a basis of gauge invariant operators 
constructed from two kinds of $u(N)$ matrices \footnote{
The paper \cite{0709.2158} originally 
studied gauge invariant operators built from 
$X$ and $X^{\dagger}$, 
but the construction 
can be straightforwardly applied to 
holomorphic operators built from 
two types of complex matrices 
$X$ and $Y$. 
See \cite{0910.2170} for another use of the Brauer algebra 
to deal with the global indices. 
} with the help of the Brauer algebra. 
For 
holomorphic operators constructed from $m$ $X$s and $n$ $Y$s, 
our basis is 
\begin{eqnarray}
O^{\gamma}_{A,ij}(X,Y)
\equiv
tr_{m,n}\left(Q^{\gamma}_{A,ij}
X^{\otimes m}\otimes (Y^{T})^{\otimes n}\right), 
\label{brauerbasis}
\end{eqnarray}
where $Q^{\gamma}_{A,ij}$ is given by a linear combination of 
elements in the Brauer algebra. 
In this paper we do not need an explicit form of 
$Q^{\gamma}_{A,ij}$. 
The definition of $Q^{\gamma}_{A,ij}$ and $O^{\gamma}_{A,ij}(X,Y)$, 
and the meaning of the labels are briefly summarised in 
appendix \ref{sec:rolebrauer}.
One striking property of 
the basis is the operators 
have diagonal two-point functions at classical level 
\cite{0709.2158}:
\begin{eqnarray}
\langle 
O^{\gamma}_{A,ij}[x]^{\dagger}
O^{\gamma^{\prime}}_{A^{\prime},i^{\prime}j^{\prime}}[0]
 \rangle _{0}\propto
\delta_{\gamma\gamma^{\prime}}
 \delta_{AA^{\prime}}
 \delta_{ii^{\prime}}\delta_{jj^{\prime}}
\frac{1}{x^{2(m+n)}},  
\label{orthogonaltwopoint}
\end{eqnarray}
where  $\langle \cdots \rangle _{0}$ means 
the tree level correlator. 
The construction of the basis 
was motivated by the fact that 
the chiral primary operators 
with diagonal two-point functions 
are classified by a 
Young diagram, which 
can be identified with giant gravitons \cite{0111222,0205221}. 

Researches along this line for the other sectors 
have been reported in 
\cite{0711.0176,0801.2061,0805.3025,0806.1911,0807.3696,0810.4217,0910.2170}. 
An idea of these works is to exploit 
algebras or groups which are dual to $U(N)$ in the meaning of 
Schur-Weyl duality (see \cite{0807.3696,0804.2764} 
from this point of view) to organise 
the multi-trace structure of gauge invariant operators. 

\vspace{0.4cm}

In this paper, 
we study the 
sector composed of two types of complex scalar fields $X$ and $Y$ 
in ${\cal N}=4$ SYM (the $su(2)$ sector) 
and focus on the dilatation operator including non-planar corrections, 
by making use of 
the Brauer algebra. 
This sector is closed under renormalisation in all order perturbation. 
Analysing the action of the one-loop dilatation operator on the basis, 
we find some gauge invariant operators which are vanished by 
the one-loop dilatation operator 
after simple algebraic 
manipulations of the Brauer algebra. 
This result would indicate that 
the Brauer algebra can be a useful tool at quantum level.\footnote{
See \cite{0701066,0801.2094} for studies of quantum corrections on 
the other bases where the symmetric group $S_{m+n}$ 
plays a role. 
See also \cite{0404066,0701.4166,0911.0967} for attempts to study non-planar corrections. 
}
Former studies on the quarter BPS operators are 
\cite{0303060,0109064,0109065,0301104,0307015}. 
See also the recent paper \cite{1002.2099}.

\vspace{0.4cm}
Here are our results. 
We will show that the following operators 
\begin{eqnarray}
O^{\gamma}(X,Y)\equiv tr_{m,n}
(P^{\gamma}X^{\otimes m}\otimes (Y^{T})^{\otimes n})
\label{operatorfromcentral}
\end{eqnarray}
satisfy 
\begin{eqnarray}
\hat{D}_{2}O^{\gamma}(X,Y)=0, 
\end{eqnarray}
where $\hat{D}_{2}$ is the one-loop dilatation operator. 
This result is valid for any $m$ and $n$. 
$P^{\gamma}$ is 
the projector associated with the irreducible representation $\gamma$ 
of the Brauer algebra, which is given by
\begin{eqnarray}
P^{\gamma}
=\sum_{A,i}Q^{\gamma}_{A,ii}. 
\end{eqnarray}
The irreducible representation of the Brauer algebra 
may be specified by the following set 
\begin{eqnarray}
\gamma=(k,\gamma_{+},\gamma_{-}), 
\end{eqnarray}
where $k$ is an integer satisfying $0 \le k \le min(m,n)$, 
$\gamma_{+}$ is a Young diagram with $m-k$ boxes and 
$\gamma_{-}$ is a Young diagram with $n-k$ boxes.  
When $N$ is finite, we have the constraint 
$c_{1}(\gamma_{+})+c_{1}(\gamma_{-}) \le N$, 
where 
$c_{1}$ denotes the length of the first column of the Young diagram. 
It follows from (\ref{orthogonaltwopoint}) 
that two-point functions are orthogonal at one-loop: 
\begin{eqnarray}
\langle 
O^{\gamma}[x]^{\dagger}
O^{\gamma^{\prime}}[0]
 \rangle _{1}\propto
\delta_{\gamma\gamma^{\prime}}
\frac{1}{x^{2(m+n)}}. 
\end{eqnarray}

\vspace{0.2cm}

Our way of finding eigenstates of the dilatation operator is based on 
a simple algebraic argument. In particular  
what we need to prove our claim 
is only the property that 
$P^{\gamma}$ are projection operators associated with 
irreducible representations of the Brauer algebra. 
We have not disclosed the operator mixing completely, but 
the results 
in this paper may be interpreted as a message that 
the classification of operators 
in terms of Brauer algebras or symmetric groups could be 
a promising way to deal with 
full non-planar corrections.

\vspace{0.2cm}

The structure of this paper is as follows. 
In Section \ref{sec:dilatationaction}, 
we derive
the action of the one-loop dilatation operator 
in terms of the Brauer algebra. 
Section \ref{sec:centralprojector} 
will be given to the proof of the claim 
that $O^{\gamma}(X,Y)$ may be quarter BPS operators at one-loop. 
The operators are classified by an integer $k$. 
We will study the sectors which are labelled by 
$k=0$ and $k=m=n$ 
in Section \ref{sec:somesectors}. 
We shall provide a brief explanation on the role of the Brauer algebra 
and the mathematical meaning of the operators $P^{\gamma}$ and $Q^{\gamma}_{A,ij}$ in 
Appendix \ref{sec:rolebrauer}. 
The mixing of the basis under the dilatation operator will be discussed in 
Appendix \ref{sec:mixingbasis}.


\section{One-loop dilatation operator on the Brauer basis}
\label{sec:dilatationaction}
\hspace{0.46cm}
In this section, we study the action of the 
one-loop 
dilatation generator on the basis composed by the Brauer algebra. 

In perturbation theory, the dilatation generator 
can be expanded in power series of the coupling constant as 
\begin{eqnarray}
\hat{D}=\sum_{l=0}\left(
\frac{g_{YM}^{2}}{16\pi^{2}}\right)^{l}\hat{D}_{2l}, 
\end{eqnarray}
where 
$\hat{D}_{2l}$ is the $l$-loop dilatation generator. 
For the $su(2)$ sector, the concrete form has been known 
\cite{0303060,0208178,0212269} (see \cite{0407277} for a review):
\begin{eqnarray}
&&
\hat{D}_{0}=tr(X\check{X}+Y\check{Y}),
\cr
&&
\hat{D}_{2}=-2tr([X,Y][\check{X},\check{Y}])\equiv -2\hat{H}, 
\end{eqnarray}
where $\check{X}$ is the derivative, and 
when $U(N)$ is the gauge group, which is the case of this paper, 
it acts as 
$(\check{X})_{ij} X_{kl}=\delta_{il}\delta_{jk}$.

\begin{figure}[t]
\begin{center}
 \resizebox{!}{5cm}{\includegraphics{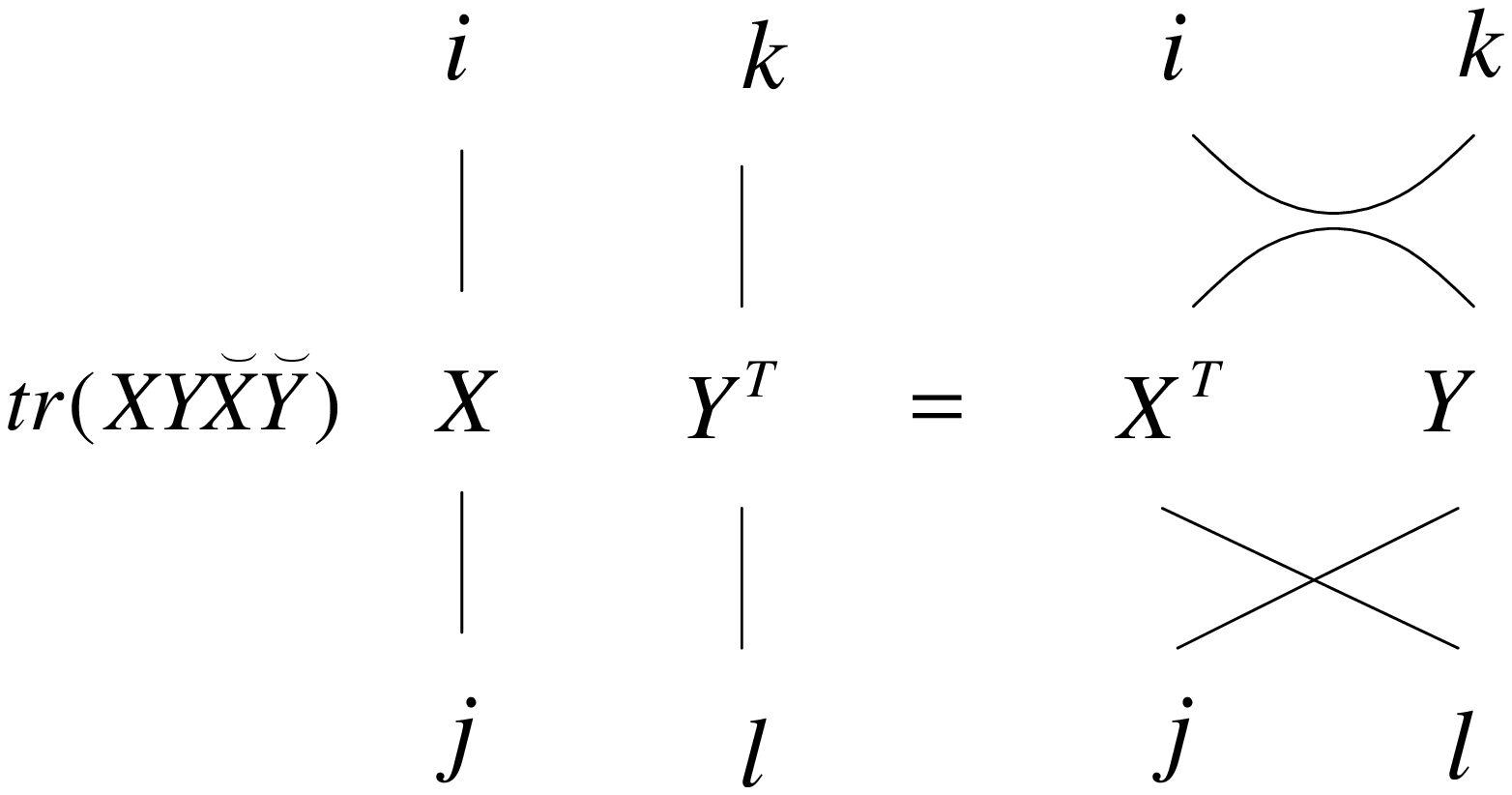}}
\caption{The action of $tr(XY\check{X}\check{Y})$ on $X \otimes Y^{T}$.}
\label{fig:dilatation_action}
\end{center}
\end{figure}

First consider the action of $\hat{H}$ on $X_{ij}(Y^{T})_{kl}$: 
\begin{eqnarray}
\hat{H} X_{ij}(Y^{T})_{kl}
&=&\left(
([X,Y])_{mn}(\check{X})_{no}(\check{Y})_{om}
-
([X,Y])_{mn}(\check{Y})_{no}(\check{X})_{om}
\right)X_{ij}(Y^{T})_{kl}
\cr
&=&
([X,Y])_{lj}\delta_{ik}
-
([X,Y])_{ik}\delta_{jl}
\cr
&=&
X_{lm}(Y^{T})_{jm}\delta_{ik}
-
X_{mj}(Y^{T})_{ml}\delta_{ik}
-
X_{im}(Y^{T})_{km}\delta_{jl}
+
X_{mk}(Y^{T})_{mi}\delta_{jl}.
\end{eqnarray}
The first term is depicted in Figure \ref{fig:dilatation_action}. 
Because the derivatives act via the Leibniz rule,  
the action of $\hat{H}$ on the basis 
(\ref{brauerbasis})
can be expressed as follows: 
\begin{eqnarray}
\hat{H} O^{\gamma}_{A,ij}&=&
\sum_{r,s}tr_{m,n}(
\sigma_{r,s} 
Q^{\gamma}_{A,ij}C_{r,s}
{\bf T}_{r}X^{\otimes m}
\otimes 
{\bf T}_{s}(Y^{T})^{\otimes n})
\cr
&&
-
\sum_{r,s}
tr_{m,n}(Q^{\gamma}_{A,ij}C_{r,s} X^{\otimes m}\otimes (Y^{T})^{\otimes n})
\cr
&&
-
\sum_{r,s}
tr_{m,n}(C_{r,s} Q^{\gamma}_{A,ij}X^{\otimes m}\otimes (Y^{T})^{\otimes n})
\cr
&&
+
\sum_{r,s}tr_{m,n}(
C_{r,s}
Q^{\gamma}_{A,ij}\sigma_{r,s} 
{\bf T}_{r}X^{\otimes m}
\otimes 
{\bf T}_{s}(Y^{T})^{\otimes n}),
\label{deltaonbrauerbasis}
\end{eqnarray}
where we have introduced the operation ${\bf T}_{i}$ 
to take the transpose on 
the $i$-th operator as 
\begin{eqnarray}
&&
{\bf T}_{r}X^{\otimes m}=
X^{\otimes r-1}\otimes X^{T}\otimes X^{m-r}
\cr
&&
{\bf T}_{s}(Y^{T})^{\otimes n}=
(Y^{T})^{\otimes s-1}\otimes Y\otimes (Y^{T})^{n-s}. 
\end{eqnarray}
$C_{r,s}$ is an element in the Brauer algebra, 
which connects the $r$-th element of $X^{\otimes m}$ with 
$s$-th element of $(Y^{T})^{\otimes n}$ 
($1\le r \le m$, $1\le s \le n$), and 
is expressed in the upper part of $X^{T}$ and $Y$  
in Figure \ref{fig:dilatation_action}. 
$\sigma_{r,s} $ is a permutation acting on 
the $r$-th element of $X^{\otimes m}$ and 
the $s$-th element of $(Y^{T})^{\otimes n}$, 
which is represented as a cross 
in Figure \ref{fig:dilatation_action}. 
Note that $\sigma_{r,s}$ is not an element in the Brauer algebra. 

One might be worried about the gauge invariance of the first term and 
forth term because $X^{T}$, $Y$ appear instead of 
$X$, $Y^{T}$. 
However, 
the gauge invariance is kept consistent 
because of the existence of $\sigma_{r,s} $. 
In order to see the gauge invariance more manifestly, 
it will be helpful to have another expression of those terms. 
The first term of (\ref{deltaonbrauerbasis}) 
may be rewritten as 
\begin{eqnarray}
tr_{m,n}(
\sigma_{r,s} 
Q^{\gamma}_{A,ij}C_{r,s}
{\bf T}_{r}X^{\otimes m}
\otimes 
{\bf T}_{s}(Y^{T})^{\otimes n})
=
tr_{m,n}(
Q^{\gamma}_{A,ij}C_{r,s}
{\bf P}_{r,s}X^{\otimes m}
\otimes 
(Y^{T})^{\otimes n}),
\end{eqnarray}
where we have introduced the operation ${\bf P}_{r,s}$ 
to exchange 
the $r$-th $X$ with the $s$-th $Y$ 
\begin{eqnarray}
&&
{\bf P}_{r,s}X^{\otimes m}
\otimes (Y^{T})^{\otimes n}=
X^{\otimes r-1}\otimes Y \otimes X^{m-r}
\otimes 
(Y^{T})^{\otimes s-1}\otimes X^{T}\otimes (Y^{T})^{n-s}. 
\end{eqnarray}
Thus we have 
\footnote{
A similar equation 
was found in unpublished work of 
T.~Brown and S.~Ramgoolam (Oct. 2008). 
}
\begin{eqnarray}
\hat{H} O^{\gamma}_{A,ij}&=&
\sum_{r,s}tr_{m,n}(
Q^{\gamma}_{A,ij}C_{r,s}
{\bf P}_{r,s}X^{\otimes m}
\otimes (Y^{T})^{\otimes n})
\cr
&&
-
\sum_{r,s}
tr_{m,n}(Q^{\gamma}_{A,ij}C_{r,s} X^{\otimes m}\otimes (Y^{T})^{\otimes n})
\cr
&&
-
\sum_{r,s}
tr_{m,n}(C_{r,s} Q^{\gamma}_{A,ij}X^{\otimes m}\otimes (Y^{T})^{\otimes n})
\cr
&&
+
\sum_{r,s}tr_{m,n}(
C_{r,s}
Q^{\gamma}_{A,ij}
{\bf P}_{r,s}X^{\otimes m}
\otimes (Y^{T})^{\otimes n}).
\label{deltaonbrauerbasis2}
\end{eqnarray}
In the equation, we have neither the transpose
nor $\sigma_{r,s}$.  Note that the second term is the same as 
the third term because of 
$(\sum_{r,s} C_{r,s})Q^{\gamma}_{A,ij}=Q^{\gamma}_{A,ij}
(\sum_{r,s} C_{r,s}) $ 
(see Appendix \ref{sec:mixingbasis}). 

Because 
each term in (\ref{deltaonbrauerbasis}) 
(or (\ref{deltaonbrauerbasis2}))
is gauge invariant, it may be 
expressed by a linear combination of the basis $O^{\gamma}_{A,ij}$.  
This can tell us how the basis mixes 
under the action of 
the dilatation operator. 
We will discuss this point in Appendix \ref{sec:mixingbasis}.


\section{Quarter BPS operators and central projectors}
\label{sec:centralprojector}
\hspace{0.46cm}
In this section, we shall prove our claim that 
the gauge invariant operators (\ref{operatorfromcentral}) 
are 
in the kernel of the one-loop dilatation operator.  
In this proof the only essential element is 
$P^{\gamma}$ to be central elements in the Brauer algebra. 
In particular  
they commute with the contractions:
\begin{eqnarray}
C_{r,s}P^{\gamma}=P^{\gamma}C_{r,s}. 
\label{centraloperator}
\end{eqnarray} 
Hence, for 
$O^{\gamma}=tr_{m,n}(P^{\gamma}X^{\otimes m }\otimes Y^{T\otimes n})$, 
we have the following action of the dilatation operator 
\begin{eqnarray}
\hat{H} O^{\gamma}
&=&
2\sum_{r,s}tr_{m,n}( 
C_{r,s}P^{\gamma}
{\bf P}_{r,s}X^{\otimes m}
\otimes (Y^{T})^{\otimes n})
\cr
&&
-
2\sum_{r,s}tr_{m,n}(C_{r,s} 
P^{\gamma}X^{\otimes m}\otimes (Y^{T})^{\otimes n}).
\label{dilatationoncentralprojector}
\end{eqnarray}
We shall see a cancellation between the first term and the second term. 

\begin{figure}[t]
\begin{center}
 \resizebox{!}{4.5cm}{\includegraphics{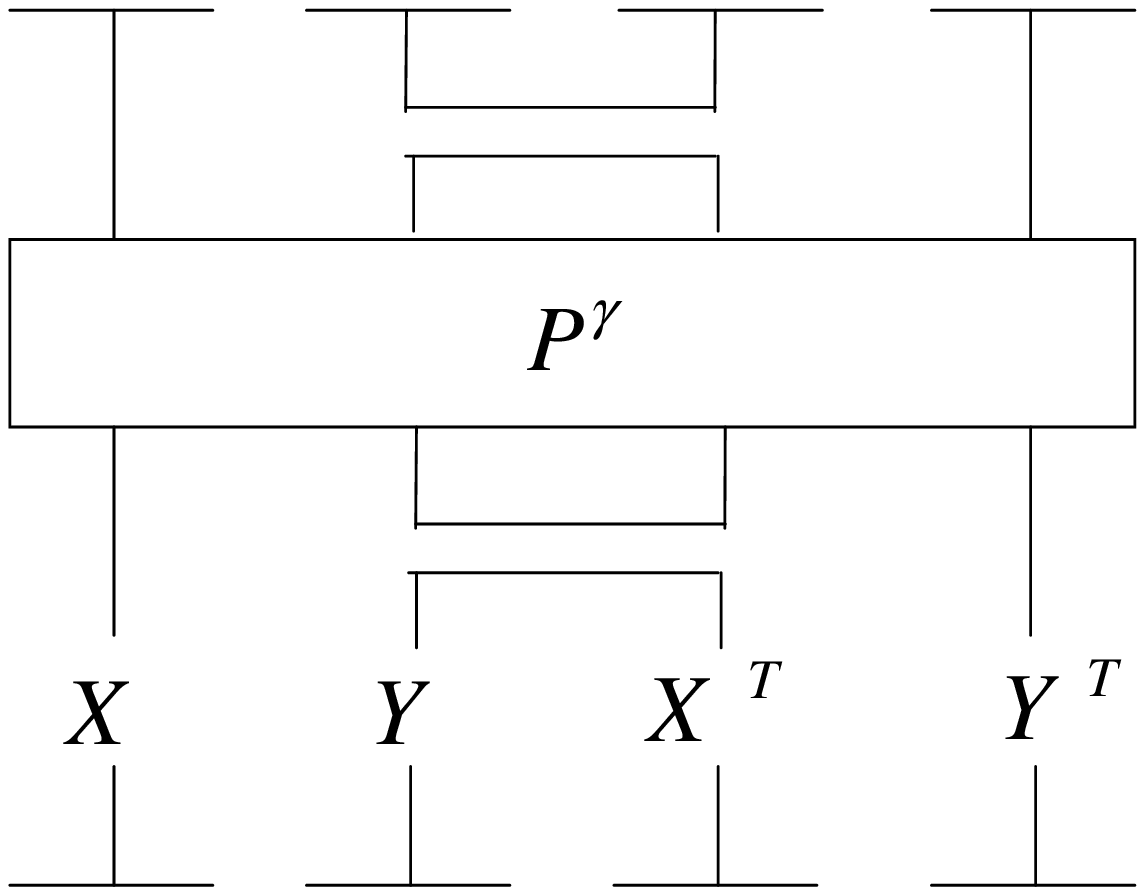}}
\caption{A diagrammatic representation of the second line 
in (\ref{firsttermcalculation}) 
at $m=n=2, r=2, s=1$. 
The top lines and the bottom lines are identified 
to express a trace $tr_{2,2}$. 
For more on the diagrammatic representation, 
see \cite{0709.2158,0205221}.}
\label{secondlinefigure}
\end{center}
\end{figure}

The equation 
(\ref{centraloperator}) 
allows us to have the following equation 
(note that $C_{r,s}^{2}=NC_{r,s}$)
\begin{eqnarray}
C_{r,s}P^{\gamma}=\frac{1}{N}
C_{r,s}P^{\gamma}C_{r,s}. 
\end{eqnarray}
Making use of this equation, the first term in 
(\ref{dilatationoncentralprojector}) can be
\begin{eqnarray}
&&
tr_{m,n}( 
C_{r,s}P^{\gamma}
{\bf P}_{r,s}X^{\otimes m}
\otimes (Y^{T})^{\otimes n})
\cr
&=&
\frac{1}{N}
tr_{m,n}( 
C_{r,s}P^{\gamma}C_{r,s}
{\bf P}_{r,s}X^{\otimes m}
\otimes 
(Y^{T})^{\otimes n})
\cr
&=&
\frac{1}{N}
tr(XY)
tr_{m,n}( 
C_{r,s}P^{\gamma}
{\bf I}_{r}X^{\otimes m}
\otimes 
{\bf I}_{s}(Y^{T})^{\otimes n}),
\label{firsttermcalculation}
\end{eqnarray}
where we have introduced the operation ${\bf I}_{i}$ to replace 
the $i$-th matrix with the identity:
\begin{eqnarray}
&&
{\bf I}_{r}X^{\otimes m}=
X^{\otimes r-1}\otimes 1\otimes X^{m-r}
\cr
&&
{\bf I}_{s}(Y^{T})^{\otimes n}=
(Y^{T})^{\otimes s-1}\otimes 1\otimes (Y^{T})^{n-s}. 
\end{eqnarray}
A diagrammatic representation of the second term in 
(\ref{firsttermcalculation}) 
is provided in Figure 
\ref{secondlinefigure}. 
Similarly the second term in 
(\ref{dilatationoncentralprojector})
may be rewritten as
\begin{eqnarray}
&&
tr_{m,n}( 
C_{r,s}P^{\gamma}
X^{\otimes m}
\otimes 
(Y^{T})^{\otimes n})
\cr
&=&
\frac{1}{N}
tr_{m,n}( 
C_{r,s}P^{\gamma}C_{r,s}
X^{\otimes m}
\otimes 
(Y^{T})^{\otimes n})
\cr
&=&
\frac{1}{N}
tr(XY)
tr_{m,n}( 
C_{r,s}P^{\gamma}
{\bf I}_{r}X^{\otimes m}
\otimes 
{\bf I}_{s}(Y^{T})^{\otimes n}). 
\end{eqnarray}
Hence we conclude that 
\begin{eqnarray}
\hat{H}O^{\gamma}=0. 
\end{eqnarray}
We have shown that the operators composed by 
the projectors associated with irreducible representations 
of the Brauer algebra are in the kernel of the one-loop 
dilatation operator.

\quad 

Before finishing this section, let us count the number of 
the operators. 
For a fixed $(m,n)$, 
the number of the operators is 
\begin{eqnarray}
\sum_{k=0}^{min(m,n)}p(m-k)p(n-k), 
\end{eqnarray}
where $p(m)$ is the number of partitions of $m$. 
When $N$ is finite, 
we have to impose $c_{1}(\gamma_{+})+c_{1}(\gamma_{-})\le N$. 
A list of the operators for some $(m,n)$ is shown 
in Table \ref{listcentral}. 

One may wonder whether 
the operators found here exhaust BPS operators. 
We may answer this question by consulting 
the partition function over BPS states 
\cite{
0510251,0608050,0704.1038}, and so it appears that there exist 
more 
BPS operators \footnote{
We thank 
S.~Ramgoolam and J.~Pasukonis 
for discussions on this point. 
}. 
It would be interesting to know how 
the other operators 
can be labelled by representations.

\quad

In general 
protected operators built from $X$ and $Y$ 
are classified into 
quarter BPS 
operators and ($SU(4)$ descendants of) the half BPS operators. 
(This is found by group theory. 
See \cite{0109064,0704.1038} for example.) 
Because the number of 
the half BPS operators built from $(m+n)$ $X$'s
is equivalent to the number of partitions of $m+n$, 
there can be $p(m+n)$ half BPS operators in the basis (\ref{brauerbasis}). 
We have considered only the kernel condition 
$\hat{H}O=0$, 
our operators are in general linear combinations of 
the half BPS operators and the quarter BPS operators. 
From the counting, at least some of quarter BPS operators 
are in the list of the operators we found. 
In this basis 
the global $SU(4)$ reprentation is not easy to see. 
Hence we do not have a good way to classify our operators 
based on the global representation at the moment. 
But an interesting thing to realise is the fact that  
some of quarter BPS operators may be combined with 
the other BPS operators to be labelled by irreducible representations 
of the Brauer algebra, 
which are more relevant to the construction of 
an orthogonal set.  
Complete classification of 
protected operators and non-protected operators 
based on the Brauer algebra is left as 
an important future problem.

\begin{table}[t]
\begin{center}
\begin{tabular}{l|c|r}
\hline
 $(1,1)$& $\gamma_{+}$ & $\gamma_{-}$ \\ \hline \hline
$k=0$ & $[1]$ & $[1]$ \\
$k=1$ & $\emptyset$ & $\emptyset$ \\
\hline
\end{tabular} 
\hspace{0.4cm}
\begin{tabular}{l|c|r}
\hline
 $(2,1)$& $\gamma_{+}$ & $\gamma_{-}$ \\ \hline \hline
$k=0$ & $[2]$ & $[1]$ \\
 & $[1,1]$ & $[1]$ \\
$k=1$ & $[1]$ & $\emptyset$ \\
\hline
\end{tabular}
 
\vspace{0.4cm}

\begin{tabular}{l|c|r}
\hline
 $(2,2)$& $\gamma_{+}$ & $\gamma_{-}$ \\ \hline \hline
$k=0$ & $[2]$ & $[2]$ \\
 & $[2]$ & $[1,1]$ \\
 & $[1,1]$ & $[2]$ \\
 & $[1,1]$ & $[1,1]$ \\
$k=1$ & $[1]$ & $[1]$ \\
$k=2$ & $\emptyset$ & $\emptyset$ \\
\hline
\end{tabular} 
\hspace{0.4cm}
\begin{tabular}{l|c|r}
\hline
 $(3,1)$& $\gamma_{+}$ & $\gamma_{-}$ \\ \hline \hline
$k=0$ & $[3]$ & $[1]$ \\
 & $[2,1]$ & $[1]$ \\
 & $[1,1,1]$ & $[1]$ \\
$k=1$ & $[2]$ & $\emptyset$ \\
 & $[1,1]$ & $\emptyset$ \\
\hline
\end{tabular} 
\hspace{0.4cm}
\begin{tabular}{l|c|r}
\hline
 $(3,2)$& $\gamma_{+}$ & $\gamma_{-}$ \\ \hline \hline
$k=0$ & $[3]$ & $[2]$ \\
 & $[2,1]$ & $[2]$ \\
 & $[1,1,1]$ & $[2]$ \\
  & $[3]$ & $[1,1]$ \\
   & $[2,1]$ & $[1,1]$ \\
    & $[1,1,1]$ & $[1,1]$ \\
$k=1$ & $[2]$ & $[1]$ \\
 & $[1,1]$ & $[1]$ \\
 $k=2$ & $[1]$ & $\emptyset$ \\
\hline
\end{tabular} 
\end{center}
\caption{ List of operators for 
$(m,n)=(1,1)$, $(2,1)$, $(2,2)$, $(3,1)$, $(3,2)$. 
If $N$ is finite, we have $c_{1}(\gamma_{+})+c_{1}(\gamma_{-})\le N$. }
\label{listcentral}
\end{table}


\section{Characteristic sectors}
\label{sec:somesectors}
\hspace{0.46cm}
Our operators are labelled by irreducible representations of 
the Brauer algebra, which are determined by the set 
$\gamma=(k,\gamma_{+},\gamma_{-})$. 
It is interesting to recognise that they are classified by an integer $k$. 
In this section, we will have a closer look on two interesting 
sectors: $k=0$ and $k=m=n$. 


\subsection{$k=0$}
\label{sec:k=0sector}
\hspace{0.46cm}
The sector labelled by $k=0$ is characterised by 
the interesting equation 
\begin{eqnarray}
C_{r,s}P^{\gamma (k=0)}=0, 
\label{cp=0}
\end{eqnarray}
for any ($r,s$). 
\footnote{
This equation was 
exploited to provide concrete forms of 
$k=0$ projectors in \cite{0709.2158}. 
}
Hence 
we immediately conclude 
from (\ref{dilatationoncentralprojector}) that  
$\hat{H} O^{\gamma (k=0)}=0$. 

We also find out that 
this class does not receive quantum corrections 
at two-loop, using the equation (\ref{cp=0}). 
The dilatation operator at two-loop was given in 
\cite{0303060}, where each term contains 
$(\check{X}\check{Y})_{ij}$ or $(\check{Y}\check{X})_{ij}$. 
The action of them on $X\otimes Y^{T}$ is 
\begin{eqnarray}
&&
(\check{X}\check{Y})_{pq} X_{ij}(Y^{T})_{kl}
=\delta_{ik}\delta_{pj}\delta_{ql},
\cr
&&
(\check{Y}\check{X})_{pq} X_{ij}(Y^{T})_{kl}
=\delta_{jl}\delta_{pk}\delta_{qi}. 
\end{eqnarray}
The appearance of $\delta_{ik}$ in the first line 
and $\delta_{jl}$ in the second line 
has the same effect as the contraction $C$ on 
$P^{\gamma (k=0)}$  
when the action on $O^{\gamma (k=0)}$ is considered. 
Therefore the equation (\ref{cp=0}) 
means that 
the $k=0$ operators 
are annihilated by the dilatation operator at two-loop.

One unique property of this sector is that 
the $k=0$ operators do not appear in the image of $\hat{D}_{2}$. 
The leading term of the $k=0$ operators is the product of 
a Schur polynomial of $X$ and a Schur polynomial of $Y$ \cite{0709.2158}. 
On the other hand, the dilatation operator always combine 
an $X$ and a $Y$ in a trace because of the contraction. 
Therefore we can not get the leading term of the 
$k=0$ operators as an image of the dilatation operator. 
This indicates that 
no mixing happens between 
$k=0$ operators and $k\neq 0$ operators. 

\quad 

The $k=0$ operators have the interesting form:
\begin{eqnarray}
&&
tr_{m,n}(P^{(k=0,\gamma_{+},\gamma_{-})}
X^{\otimes m}\otimes Y^{T \otimes n})
\cr
&=&
Dim\gamma_{+}\gamma_{-}\frac{m!n!}{d_{\gamma_{+}}d_{\gamma_{-}}}
\frac{1}{N^{m+n}}
tr_{m,n}\left(\Omega_{m+n}^{-1}p_{\gamma_{+}}p_{\gamma_{-}}
X^{\otimes m}\otimes Y^{\otimes n}\right). 
\end{eqnarray}
This was obtained in \cite{0709.2158} 
(see (4.7) and (4.8) in the paper). 
In this expression, 
$\Omega_{m+n}^{-1}$ is the omega factor considered 
in the symmetric group $S_{m+n}$, and 
$p_{\gamma_{+}}$ and $p_{\gamma_{-}}$ 
are projection operators in $S_{m}$ and $S_{n}$. 

The $k=0$ operators for some $(m,n)$ are shown explicitly 
for the $X$-$X^{\dagger}$ sector 
in \cite{0709.2158}. 
The $X$-$Y$ system is obtained 
by replacing $X^{\dagger}$ with $Y$. 

\vspace{0.4cm}

Let me mention 
on the role of (\ref{cp=0}) for  
non-holomorphic operators constructed from $X$. 
A $k=0$ operator is a linear combination of single traces and multi-traces. 
An interesting property is that 
divergences arising 
from self-contractions cancel among those terms due 
to the equation (\ref{cp=0}) 
\cite{0709.2158,0910.2170}. 
One may keep in mind that the equation (\ref{cp=0}) 
plays interesting and different 
roles in the two systems ($X$-$Y$ and $X$-$X^{\dagger}$). 
See also \cite{0911.4408} for physics related to the $k=0$ sector.

\subsection{$k=m=n$}
\hspace{0.46cm}
In this subsection we study the operators  
at $k=m=n$. This sector was investigated  
in \cite{0911.4408}.  
We leave some details 
(the definition of $C_{(k)}$, the derivation of (\ref{k=mprojectorR}) and 
how to rewrite the first line to get the last expression in (\ref{k=mcentral}))
to \cite{0911.4408} 
because the purpose of this subsection is not to review 
the calculations given in it. 

When $k=m=n$, some labels of $Q^{\gamma}_{A,ij}$ are trivial, thereby  
$Q^{\gamma}_{A,ij}$ 
are labelled by a single Young diagram 
$\alpha$ with $k$ boxes (we define $P^{\gamma(k=m)}_{\alpha}\equiv Q^{\gamma(k=m)}_{\alpha}$): 
\begin{eqnarray}
P_{\alpha}^{\gamma(k=m)}=\frac{d_{\alpha}}{k!Dim\alpha}C_{(k)}p_{\alpha}. 
\label{k=mprojectorR}
\end{eqnarray} 
The central projector $P^{\gamma (k=m)}$ can be obtained by the sum of 
all projectors in this sector
\footnote{
In this sector, there is only one central projector because 
$\gamma_{+}=\gamma_{-}=\emptyset$. 
} 
\begin{eqnarray}
P^{\gamma(k=m)}=\sum_{\alpha\vdash k}P_{\alpha}^{\gamma(k=m)}
=\frac{1}{N^{k}}\Omega_{k}^{-1}C_{(k)}, 
\end{eqnarray}
where we have used 
the formula for the inverse of the omega factor:
\begin{eqnarray}
\Omega_{k}^{-1}
=\frac{N^{k}}{k!}\sum_{\alpha\vdash k}
\frac{d_{\alpha}}{Dim\alpha}p_{\alpha}.
\end{eqnarray}
The omega factor played an important role in the large $N$ expansion of 
two-dimensional Yang-Mills \cite{9303046,9411210}.

Gauge invariant operators 
relevant to the projectors can be written down as  
\begin{eqnarray}
O^{\gamma (k=m)}(X,Y)
&=&
tr_{k,k}(P^{\gamma}X^{\otimes k}\otimes (Y^{T})^{\otimes k})
\cr
&=&
\frac{1}{N^{k}}
tr_{k}(\Omega_{k}^{-1}S^{\otimes k}),
\label{k=mcentral}
\end{eqnarray}
where we have defined $S=XY$. 
This sector may be characterised by the fact that 
these operators are invariant under 
$X\rightarrow gX$, $Y\rightarrow Yg^{-1}$ 
or $X\rightarrow Xg$, $Y\rightarrow g^{-1}Y$. 
These transformations are introduced in 
\cite{0807.3696} to measure labels of orthogonal sets. 
More remarks are in \cite{0911.4408}. 

For concreteness, we present explicit forms:
\begin{eqnarray}
O^{\gamma(k=2)}
=
\frac{1}{N(N^{2}-1)}\left(
N(trS)^{2}-tr(S^{2})
\right)
\end{eqnarray}
for $m=n=2$, and 
\begin{eqnarray}
O^{\gamma(k=3)}
=
\frac{1}{N(N^{2}-1)(N^{2}-4)}
\left(
6(N^{2}-2)(trS)^{3}-18N(trS)tr(S^{2})+24tr(S^{3})
\right)
\end{eqnarray}
for $m=n=3$. 


\vspace{1.2cm}

\noindent 
{\bf Acknowledgements}

\vspace{0.2cm}

I would like to thank 
Sanjaye Ramgoolam for helpful discussions and for 
reading drafts of the paper. 
I also thank Tom Brown, 
Robert de Mello Koch, 
Maria Pilar Garcia del Moral, 
Yasuyuki Hatsuda, 
Paul Heslop,  
Jurgis Pasukonis, David Turton 
for stimulating discussions. 
I thank theoretical physics laboratory in riken for 
the opportunity to stay. 
This work has been partially supported by the Research Grants MICINN FPA2009-07122, 
IB09-069 PCTI Asturias 2006-2009, FEDER 2007-2013, and MEC-DGI CSD2007-00042 (Spain).

\vspace{1.2cm}

\renewcommand{\theequation}{\Alph{section}.\arabic{equation}}
\appendix 

\section{On the role of the Brauer algebra}
\label{sec:rolebrauer}
\setcounter{equation}{0} 
\hspace{0.46cm}
In this section, we shall make an brief description on 
the role of the Brauer algebra in constructing a set of gauge invariant operators,  
with emphasis on group theoretic structure. 
A more useful review of Brauer algebras may be found in \cite{0709.2158}, 
and references therein. 

An $N\times N$ matrix $X$ can be viewed as 
an endomorphism acting on 
an $N$-dimensional vector space $V$, i.e. $X$ : $V \rightarrow V$. 
The tensor product $X^{\otimes n}=X\otimes \cdots \otimes X$ acts on 
$V^{\otimes n}$. 
The symmetric group $S_{n}$ can be introduced  
as a tool to organise both single trace and multi-trace.
We define the action of the symmetric group $S_{n}$
as the permutations of $n$ vector spaces.   
The tensor product space can be decomposed into irreducible representations as 
\begin{eqnarray}
V^{\otimes n}=\bigoplus_{R}
V_{R}^{U(N)}\otimes V_{R}^{S_{n}}.  
\end{eqnarray}
This is a consequence of the fact that 
the $U(N)$ action and the symmetric group action commute each other 
on the space $V^{\otimes n}$ [Schur-Weyl duality]. 
The sum is taken for 
all irreducible representations with 
$n$ boxes satisfying $c_{1}(R)\le N$, 
where $c_{1}(R)$ is the length of the first column of $R$. 
The projection operator $p_{R}$ 
associated with an irreducible representation $R$ 
can be introduced as  
an element 
in the group algebra of 
the symmetric group $S_{n}$. 
The operators defined by $tr_{n}(p_{R}X^{\otimes n})$ were shown to 
form a complete set of gauge invariant operators 
in the chiral primary sector \cite{0111222}. 
The trace $tr_{n}$ is taken in $V^{\otimes n}$. 

\vspace{0.4cm}

Let us next consider 
the space $V^{\otimes m}\otimes \bar{V}^{\otimes n}$ by 
including the complex conjugated space $\bar{V}$. 
Elements of 
the symmetric group $S_{m}\times S_{n}$ 
can act on the space as linear maps. 
The contraction $C$ can also be introduced as a map from $V\times \bar{V}$ to itself. 
(It acts as $Cv_{i}\bar{v}_{j}=\delta_{ij}\sum_{k}v_{k}\bar{v}_{k}$ for components.) 
An algebra formed by the group algebra of the symmetric group 
$S_{m}\times S_{n}$ and the contractions 
is the Brauer algebra. We shall denote it by $B_{N}(m,n)$.  
Note that the Brauer algebra is sensitive to $N$ while the symmetric group is not.
Schur-Weyl duality relevant to this case is 
\begin{eqnarray}
V^{\otimes m}\otimes \bar{V}^{\otimes n}
=\bigoplus_{\gamma}
V_{\gamma}^{U(N)}\otimes V_{\gamma}^{B_{N}(m,n)}. 
\end{eqnarray}
This gives the decomposition of the tensor product space in terms of 
irreducible representations of $U(N)$ and $B_{N}(m,n)$. 
We have the projection operator 
$P^{\gamma}$ 
associated with an irreducible representation $\gamma$ which 
is a linear combination of elements in the Brauer algebra. 
Note that it is 
in the centre of the Brauer algebra, i.e. 
\begin{eqnarray}
bP^{\gamma}=P^{\gamma}b, 
\end{eqnarray}
where $b$ is any element in $B_{N}(m,n)$. 
The space $V_{\gamma}^{B_{N}(m,n)}$ can be further decomposed into the group algebra 
of $S_{m}\times S_{n}$, which we denote by $\mathbb C(S_{m}\times S_{n})$, 
as 
\begin{eqnarray}
V_{\gamma}^{B_{N}(m,n)}=
\bigoplus_{A}V_{\gamma \rightarrow A}\otimes V_{A}^{\mathbb C(S_{m}\times S_{n})}, 
\end{eqnarray}
where $A$ runs over irreducible representations of 
the symmetric group $S_{m}\times S_{n}$, and 
$V_{\gamma \rightarrow A}$ represents the space of the multiplicity relevant to this decomposition. 
We introduce the operator $Q^{\gamma}_{A,ij}$ which acts on the space 
$V_{A}^{S_{m}\times S_{n}}$ inside $V_{\gamma}^{B_{N}(m,n)}$. 
The indices $i$,$j$ run over the space of the multiplicity, and besides, 
they behave like matrix indices as 
\begin{eqnarray}
Q^{\gamma}_{A,ij}
Q^{\gamma^{\prime}}_{A^{\prime},i^{\prime}j^{\prime}}
=\delta^{\gamma \gamma^{\prime}}
\delta_{A A^{\prime}}
\delta_{j i^{\prime}}
Q^{\gamma}_{A,i j^{\prime}}. 
\label{producQ}
\end{eqnarray}
The relation between $P^{\gamma}$ and $Q^{\gamma}_{A,ij}$ is 
\begin{eqnarray}
P^{\gamma}=\sum_{A,i}Q^{\gamma}_{A,ii}. 
\end{eqnarray}
The readers who are interested in the explicit form of $Q^{\gamma}_{A,ij}$ 
would be 
recommended to see \cite{0807.3696}.  
Acting with $Q^{\gamma}_{A,ij}$ on $X^{\otimes m}\otimes Y^{T\otimes n}$ 
and taking a trace in $V^{\otimes m}\otimes \bar{V}^{\otimes n}$, 
we get 
$tr_{m,n}\left(Q^{\gamma}_{A,ij}X^{\otimes m}\otimes Y^{T \otimes n}\right)$, 
which is the basis proposed in \cite{0709.2158}.


\section{Mixing of the basis}
\label{sec:mixingbasis}
\setcounter{equation}{0} 
\hspace{0.46cm}
In order to realise how operators mix under the action of 
the dilatation operator, 
it will be helpful to rewrite 
each term of 
(\ref{deltaonbrauerbasis}) 
in terms of the basis. 
In this section, we perform it for the third term. 

Introduce the sum of contractions,  
\begin{eqnarray}
C\equiv \sum_{r,s}C_{r,s}
=\frac{1}{(m-1)!(n-1)!}\sum_{h\in S_{m}\times S_{n}}
hC_{1,1}h^{-1}. 
\end{eqnarray}
We can easily show 
that it commutes with any element in the symmetric group $S_{m}\times S_{n}$: 
\begin{eqnarray}
C \tau = \tau C, \quad \tau \in S_{m}\times S_{n}. 
\end{eqnarray}
Such an element can be 
expressed by a linear combination of $Q^{\gamma}_{A,ij}$, 
and 
the formula was given in appendix B.2 of \cite{0807.3696}:
\begin{eqnarray}
C=\sum_{\gamma,A,ij}\frac{1}{d_{A}}\chi^{\gamma}_{A,ij}(C)
Q^{\gamma}_{A,ij}, 
\end{eqnarray}
where $d_{A}$ is the dimension of the symmetric group 
$S_{m}\times S_{n}$ associated with 
the representation $A$, and $\chi^{\gamma}_{A,ij}$ is the restricted character. 
Using this, we get 
\begin{eqnarray}
CQ^{\gamma}_{A,ij}
&=&\sum_{\gamma^{\prime},A^{\prime},i^{\prime}j^{\prime}}
\frac{1}{d_{A^{\prime}}}
\chi^{\gamma^{\prime}}_{A^{\prime},i^{\prime}j^{\prime}}(C)
Q^{\gamma^{\prime}}_{A^{\prime},i^{\prime}j^{\prime}}
Q^{\gamma}_{A,ij}
\cr
&=&\sum_{i^{\prime}}
\frac{1}{d_{A}}
\chi^{\gamma}_{A,i^{\prime}i}(C)
Q^{\gamma}_{A,i^{\prime}j}, 
\end{eqnarray}
where (\ref{producQ}) has been used.
Hence we can rewrite the third term in 
(\ref{deltaonbrauerbasis}) as 
\begin{eqnarray}
\sum_{r,s}
tr_{m,n}(C_{r,s}Q^{\gamma}_{A,ij} X^{\otimes m}\otimes (Y^{T})^{\otimes n})
=
\frac{1}{d_{A}}
\sum_{i^{\prime}}
\chi^{\gamma}_{A,i^{\prime}i}(C)
O^{\gamma}_{A,i^{\prime}j}. 
\end{eqnarray}
As far as this term is concerned, 
the mixing is strictly restricted because 
only the multiplicity index is relevant for the mixing.


\end{document}